\documentclass[useAMS]{mn2e}
\usepackage{graphicx}

\def\arcmin{\tt '}

\def\simgt{\ {\raise-.5ex\hbox{$\buildrel>\over\sim$}}\ }
\def\I{\'\i}
\def\cd{cd$^{-1}$\,}
\def\kms{kms$^{-1}$\,}

\begin{document}

\title[Asteroseismology of 12 (DD) Lacertae: photometry]
{Asteroseismology of the $\beta$ Cephei star 12 (DD) Lacertae:
photometric observations, pulsational frequency analysis and mode 
identification}
\author[G. Handler et al.]
 {G. Handler,$^{1}$ M. Jerzykiewicz,$^{2}$ E. Rodr\I guez,$^{3}$ K. 
Uytterhoeven,$^{4, 5}$ P. J. Amado,$^{3}$ \and T. N. Dorokhova,$^{6}$ N. 
I. Dorokhov,$^{6}$ E. Poretti,$^{7}$ J.-P. Sareyan,$^{8}$ L.
Parrao,$^{9}$ \and D. Lorenz,$^{1}$ D. Zsuffa,$^{10}$ R. Drummond,$^{4}$
J. Daszy{\'n}ska-Daszkiewicz,$^{2, 4}$ \and T. Verhoelst,$^{4}$ J. De
Ridder,$^{4}$ B. Acke,$^{4}$ P.-O.Bourge,$^{11}$ A. I. Movchan,$^{6}$ \and
R. Garrido,$^{3}$ M. Papar\'o,$^{10}$ T. Sahin,$^{12, 13}$ V. 
Antoci,$^{1}$ S. N. Udovichenko,$^{6}$ \and K. Csorba,$^{10}$ R.
Crowe,$^{14}$ B. Berkey,$^{14}$ S. Stewart,$^{14}$ D. Terry,$^{14}$ 
\and D. E. Mkrtichian,$^{6, 15}$ C. Aerts$^{4}$
\and \\
$^1$ Institut f\"ur Astronomie, Universit\"at Wien, T\"urkenschanzstrasse
17, A-1180 Wien, Austria\\
$^{2}$ Wroclaw University Observatory, ul. Kopernika 11, 51-622 Wroclaw, 
Poland\\
$^{3}$ Instituto de Astrofisica de Andalucia, C.S.I.C., Apdo. 3004, 18080 
Granada, Spain\\
$^{4}$ Instituut voor Sterrenkunde, K. U. Leuven, Celestijnenlaan 200B, 
B-3001 Leuven, Belgium\\
$^{5}$ Mercator Telescope, Calle Alvarez de Abreu 70, 38700 Santa Cruz de 
La Palma, Spain\\
$^{6}$ Astronomical Observatory of Odessa National University, 
Marazlievskaya, 1v, 65014 Odessa, Ukraine\\
$^{7}$ INAF-Osservatorio Astronomico di Brera, Via Bianchi 46, 23807 
Merate, Italy\\
$^{8}$ Observatoire de la C\^{o}te d'Azur, D\'epartement GEMINI - UMR 
6203, BP 4229, 06304 Nice Cedex 4, France\\
$^{9}$ Instituto de Astronomia UNAM, Apartado 70-264, C.\ U.\ 04510 Mexico 
D.\ F., Mexico\\
$^{10}$ Konkoly Observatory, Box 67, H-1525 Budapest XII, Hungary\\
$^{11}$ Institut d'Astrophysique et de G\'eophysique, Universit\'e de
Li\`ege, all\'ee du Six Ao\^{u}t 17, 4000 Li\`ege, Belgium\\
$^{12}$ Physics Department, Akdeniz University, 07058 Antalya, Turkey\\
$^{13}$ Armagh Observatory, College Hill, Armagh BT61 9DG, N. Ireland, 
UK\\
$^{14}$ Department of Physics and Astronomy, University of Hawaii - Hilo,
200 West Kawili Street, Hilo, Hawaii, 96720-4091, USA\\
$^{15}$ ARCSEC, Sejong University, Seoul, 143-747, Korea}

\date{Accepted 2005 July 17.
 Received 2005 August 13;
 in original form 2005 September 10}
\maketitle
\begin{abstract}
We report a multisite photometric campaign for the $\beta$ Cephei star 12
Lacertae. 750 hours of high-quality differential photoelectric
Str\"omgren, Johnson and Geneva time-series photometry were obtained with
9 telescopes during 190 nights. Our frequency analysis results in the
detection of 23 sinusoidal signals in the light curves. Eleven of those
correspond to independent pulsation modes, and the remainder are
combination frequencies. We find some slow aperiodic variability such as
that seemingly present in several $\beta$ Cephei stars. We perform mode
identification from our colour photometry, derive the spherical degree
$\ell$ for the five strongest modes unambiguously and provide constraints
on $\ell$ for the weaker modes. We find a mixture of modes of $0 \leq \ell
\leq 4$. In particular, we prove that the previously suspected
rotationally split triplet within the modes of 12 Lac consists of modes of
different $\ell$; their equal frequency splitting must thus be accidental.

One of the periodic signals we detected in the light curves is argued to
be a linearly stable mode excited to visible amplitude by nonlinear mode
coupling via a 2:1 resonance. We also find a low-frequency signal in the
light variations whose physical nature is unclear; it could be a parent or
daughter mode resonantly coupled. The remaining combination frequencies
are consistent with simple light-curve distortions.

The range of excited pulsation frequencies of 12 Lac may be sufficiently
large that it cannot be reproduced by standard models. We suspect that the
star has a larger metal abundance in the pulsational driving zone, a
hypothesis also capable of explaining the presence of $\beta$ Cephei stars
in the LMC.
\end{abstract}


\section{Introduction}

12 (DD) Lacertae (hereinafter briefly called 12 Lac) is one of the best
observed $\beta$ Cephei stars, a class of variable early B-type stars
whose light and radial velocity changes are due to gravity and pressure
mode pulsations of low radial order (Stankov \& Handler 2005 and
references therein). Radial velocity variations of 12 Lac were discovered
nearly one hundred years ago (Adams 1912), and light variations were
detected soon thereafter (Stebbins 1917, Guthnick 1919).

After the recognition of the pulsational nature of the light variations of 
the $\beta$ Cephei stars (Ledoux 1951) and because of the complicated 
nature of the variability of 12 Lac, the star became the target of one 
of the first worldwide observing campaigns (de Jager 1963), also called 
{\it The International Lacerta weeks}, during which a total of more than 
700 hours of time-resolved photometric and spectroscopic measurements were 
secured -- back in 1956!

The photometric measurements obtained during this prototype multisite
campaign were analysed by Barning (1963) who discovered four different
variations in the light curves (one of them spurious). Jerzykiewicz (1978) 
provided an extensive re-analysis of these and some other data and proved
the existence of five independent periodicities and one combination
frequency in the data.

High-resolution spectroscopic observations of 12 Lac, confirming the
six photometric pulsation frequencies, were carried out by Mathias et al.\ 
(1994). However, attempts to identify the underlying pulsation modes by
spectroscopic means were unsuccessful due to the complicated behaviour of
the star, and earlier mode identification results (e.g. the pioneering
work by Smith 1980) could neither be confirmed nor rejected.

The most striking feature within the pulsation frequencies of 12 Lac
is an equally spaced triplet at 5.179, 5.334 and 5.490 \cd. Owing to the
narrowness of the interval these three frequencies span, it is clear that
at least two of the underlying modes must be nonradial and it is
straightforward to speculate that all are actually components of a
rotationally split multiplet.

Because of this interesting possibility and because 12 Lac has been
for a long time the $\beta$ Cephei star with the largest number of known
pulsation modes, it is an attractive target for asteroseismic
investigations, i.e.\ deriving the interior structure of the star by
modelling its pulsation frequencies. Dziembowski \& Jerzykiewicz (1999) 
carried out such a study. They could only reproduce the equally spaced
triplet with an $\ell=2$ f-mode, and solely for models with specific
values of temperature and surface gravity. The authors suggested that
nonlinear phase-lock could provide a way out of this dilemma.

Another intriguing possibility for interpreting the pulsation spectrum of
12 Lac that was not previously realised is that the ratio of the
lowest pulsation frequency of the star (4.241 \cd) to that of another
frequency at 5.490 \cd is perfectly consistent with the expected frequency
ratio of the radial fundamental and first overtone modes. If such an
interpretation were correct, it would be rather easy to perform seismic
model computations, especially if more modes could be observationally
detected. The problem with the equally spaced triplet would vanish as well
because one of the suspected multiplet members would in fact be a radial
mode.

Clearly, the interpretation of the pulsation spectrum of 12 Lac can 
potentially be extremely rewarding. Another observational effort similar 
to {\it The International Lacerta weeks}, with modern observing methods, 
seemed therefore quite worthwhile. In particular, new observations would 
be required to provide unique mode identifications for at least the five 
known pulsation modes for seismic modelling to commence. The discovery of 
additional low-amplitude pulsations would of course allow more insights 
into the interior structure of the star as well.

As recent observing campaigns have revealed many low-amplitude pulsation
modes for some $\beta$ Cephei stars [e.g., see Aerts et al.\ (2004) for
V836 Cen, Jerzykiewicz et al.\ (2005) and references therein for $\nu$
Eridani, and Handler, Shobbrook \& Mokgwetsi (2005) for $\theta$
Ophiuchi], it only seemed logical to organise a similar effort for 12 
Lac. We have therefore carried out a multisite campaign for the star with
both photometric and spectroscopic techniques. In addition, the close-by
eclipsing binary $\beta$ Cephei star 16 (EN) Lac was observed
photometrically at the same time. Whereas we postpone the analysis of the
latter star and that of the spectroscopy to forthcoming papers, we report
here the results of the photometric measurements of 12 Lac.

\section{Observations and reductions}

Our photometric observations were carried out at nine different
observatories with small to medium-sized telescopes on three different
continents (see Table 1). In most cases, single-channel differential
photoelectric photometry was acquired; some additional CCD measurements
turned out not to be useful. Wherever possible, the Str\"omgren $uvy$
filters were used.

\begin{table*}
\caption[]{Log of the photometric measurements of 12 Lacertae. 
Observatories are ordered according to geographical longitude.}
\begin{center}
\begin{tabular}{lrcccccl}
\hline
Observatory & Longitude & Latitude & Telescope & \multicolumn{2}{c}{Amount 
of data} & Filter(s) & Observer(s)\\
& & & & Nights & h & & \\
\hline
Sierra Nevada Observatory & $-$3\degr 23\arcmin & +37\degr 04\arcmin & 0.9m 
& 18 & 81.2 & uvby & ER, PJA, RG\\
Mercator Observatory & $-$17\degr 53\arcmin & +28\degr 46\arcmin & 1.2m & 
45 & 123.9 & Geneva & KU, RD, JDD, TV\\
 & & & & & & & JDR, BA, POB\\
Fairborn Observatory & $-$110\degr 42\arcmin & +31\degr 23\arcmin & 0.75m 
APT & 55 & 201.5 & uvy & $--$\\
Lowell Observatory & $-$111\degr 40\arcmin & +35\degr 12\arcmin & 0.5m & 
19 & 97.3 & uvy & MJ\\
San Pedro Martir Observatory & $-$115\degr 28\arcmin & +31\degr 03\arcmin 
& 1.5m & 20 & 102.7 & uvby & EP, JPS, LP\\
Mt. Dushak-Erekdag Observatory & +57\degr 53\arcmin & +37\degr 55\arcmin &
0.8m & 13 & 70.7 & V & TND, NID\\
T\"ubitak National Observatory & +30\degr 20\arcmin & +36\degr 50\arcmin & 
0.5m & 1 & 2.5 & V & TS\\
Mayaki Observatory & +30\degr 17\arcmin & +46\degr 24\arcmin & 0.5m & 6 & 
13.3 & V & AIM\\
Piszk\'estet\H o Observatory & +19\degr 54\arcmin & +47\degr 55\arcmin & 
0.5m & 13 & 56.7 & V & MP, DZ, DL, VA\\
\hline
Total & & & & 190 & 749.8 \\
\hline
\end{tabular}
\end{center}
\end{table*}

However, at the Sierra Nevada (OSN) and San Pedro Martir (SPM) 
Observatories simultaneous $uvby$ photometers were available, including
the $b$ filter as well. On the other hand, the $u$ data from SPM were
unusable. At four other observatories where no Str\"omgren filters were
available we used Johnson $V$. Finally, as the photometer at the Mercator
telescope has Geneva filters installed permanently, we used this filter
system.

We chose the two ``classical'' comparison stars for 12 and 16 (EN) 
Lac: 10 Lac (O9V, $V=4.88$) and 2 And (A3Vn, $V=5.09$). Another check
star, HR 8708 (A3Vm, $V=5.81$), was additionally observed during one of 
the SPM runs.

Data reduction was begun by correcting for coincidence losses, sky
background and extinction. Whenever possible, nightly extinction
coefficients were determined with the differential Bouguer method (fitting
a straight line to a plot of differential magnitude vs.\ differential air
mass) from the measurements of the two comparison stars. Second-order
colour extinction coefficients were also determined and used to adjust the
measurements of 2 And, which suffers lower extinction because it is redder
than the other stars. On some nights the coefficients were derived from 10
Lac with the usual Bouguer method.

It turned out that 2 And is a low-amplitude $\delta$ Scuti variable. Light
variations of this star have already been strongly suspected by Sareyan et
al.\ (1997). We will discuss these variations in the forthcoming paper
devoted to 16 (EN) Lac because 2 And was in the past mostly used as the
main comparison star for the latter $\beta$ Cephei star. For the purpose
of the present work, let it suffice to say that we only found evidence for
a single periodicity in our light curves of 2 And. No evidence for
photometric variability of 10~Lac was found.

We thus proceeded by prewhitening the variability of 2 And with a fit
determined from all its differential magnitudes relative to 10 Lac from
the individual nights of measurement. The residual magnitudes of 2 And
were then combined with the 10 Lac data into a curve that was assumed to
reflect the effects of transparency and detector sensitivity changes only.
Consequently, these combined time series were binned into intervals that
would allow good compensation for the above-mentioned nonintrinsic
variations in the target star time series and were subtracted from the
measurements of 12 Lac. The binning minimises the introduction of
noise in the differential light curve of the targets.

The timings for the differential light curves were heliocentrically
corrected as the next step, and the single-colour measurements were binned
to sampling intervals similar to that of the multicolour measurements to
avoid unwanted implicit weighting effects. Finally, the photometric
zeropoints of the different instruments were compared between the
different sites and adjusted if necessary. Measurements in the Str\"omgren
$y$ and Johnson and Geneva $V$ filters were treated as equivalent due to
the same effective wavelength of these filters, and were analysed
together. This combined light curve is henceforth referred to as ``the $V$
filter data''.

The resulting final combined time series was subjected to frequency
analysis; we show some of our light curves of 12 Lac in Fig.\ 1. In
the end, we had 3239 $V$ filter measurements available (time span 197.0 d),
2301 points in Str\"omgren $v$ (time span 179.2 d), 1739 points in
Str\"omgren $u$ (time span 179.2 d) and 488 points in the Geneva filters
(time span 173.8 d)

\begin{figure*}
\includegraphics[angle=270,width=184mm,viewport=-10 5 460 750]{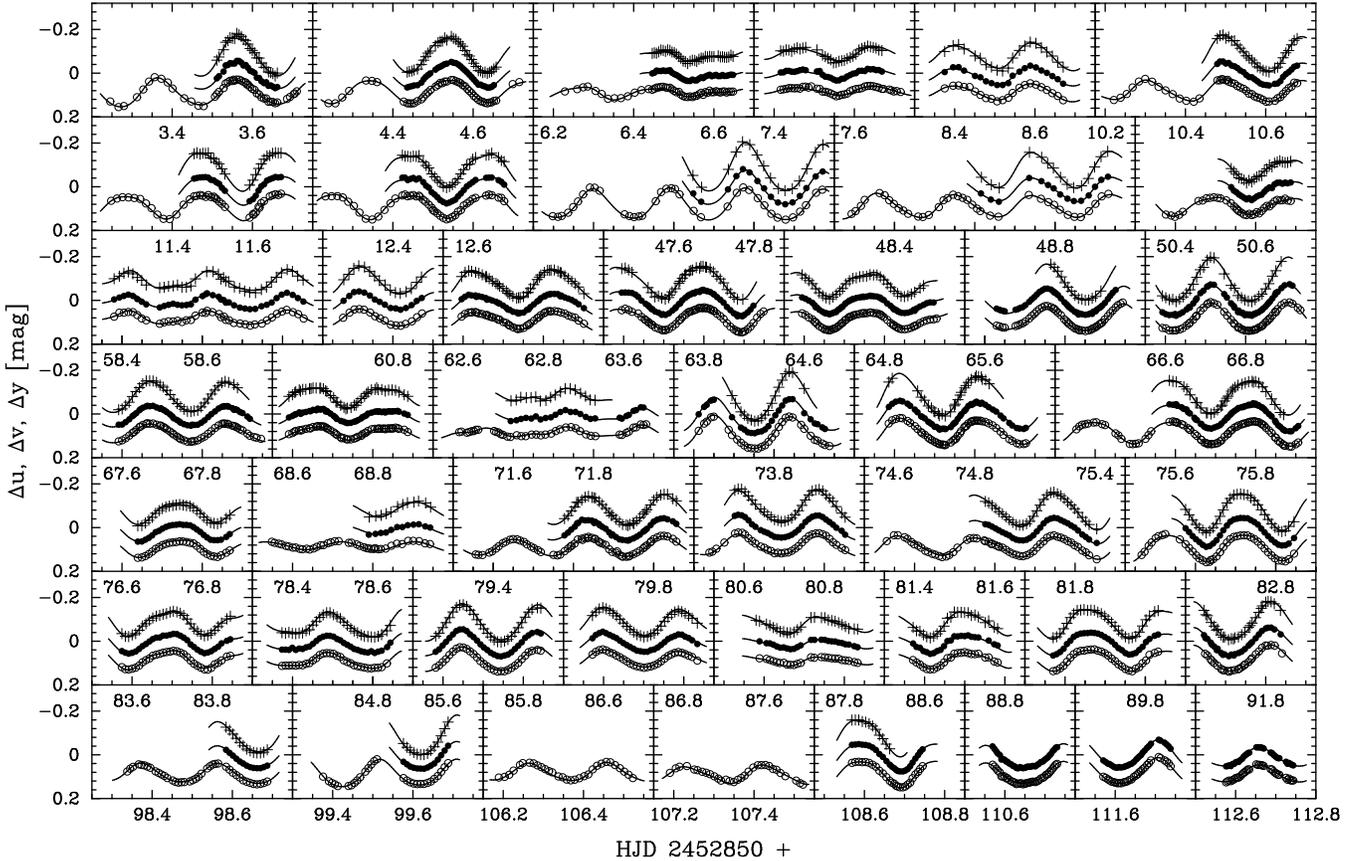}
\caption[]{Some of our observed light curves of 12 Lac. Plus signs
are data in the Str\"omgren $u$ filter, filled circles are our $v$
measurements and open circles represent the $V$ data. The full
line is a fit composed of the 23 periodicities detected in the light
curves (Table 2). The amount of data displayed is about half the total.}
\end{figure*}

\section{Frequency analysis}

Our frequency analysis was mainly performed with the program {\tt
Period98} (Sperl 1998). This package applies single-frequency power
spectrum analysis and simultaneous multi-frequency sine-wave fitting. It
also includes advanced options such as the calculation of optimal
light-curve fits for multiperiodic signals including harmonic,
combination, and equally spaced frequencies. As will be demonstrated
later, our analysis requires some of these features.

We started by computing the Fourier spectral window of the $V$ filter
data. It was calculated as the Fourier transform of a single noise-free
sinusoid with a frequency of 5.179 \cd (the strongest pulsational signal
of 12 Lac) and an amplitude of 40 mmag, sampled in the same way as
our measurements. The upper panel of Fig.\ 2 (left-hand side)
contains the result. There are some alias structures in this window
function due to our lack of measurements from Eastern Asian longitudes. We
must therefore apply caution in identifying the correct frequencies in the
light variations of the star.

\begin{figure*}
\includegraphics[width=164mm,viewport=-03 05 482 490]{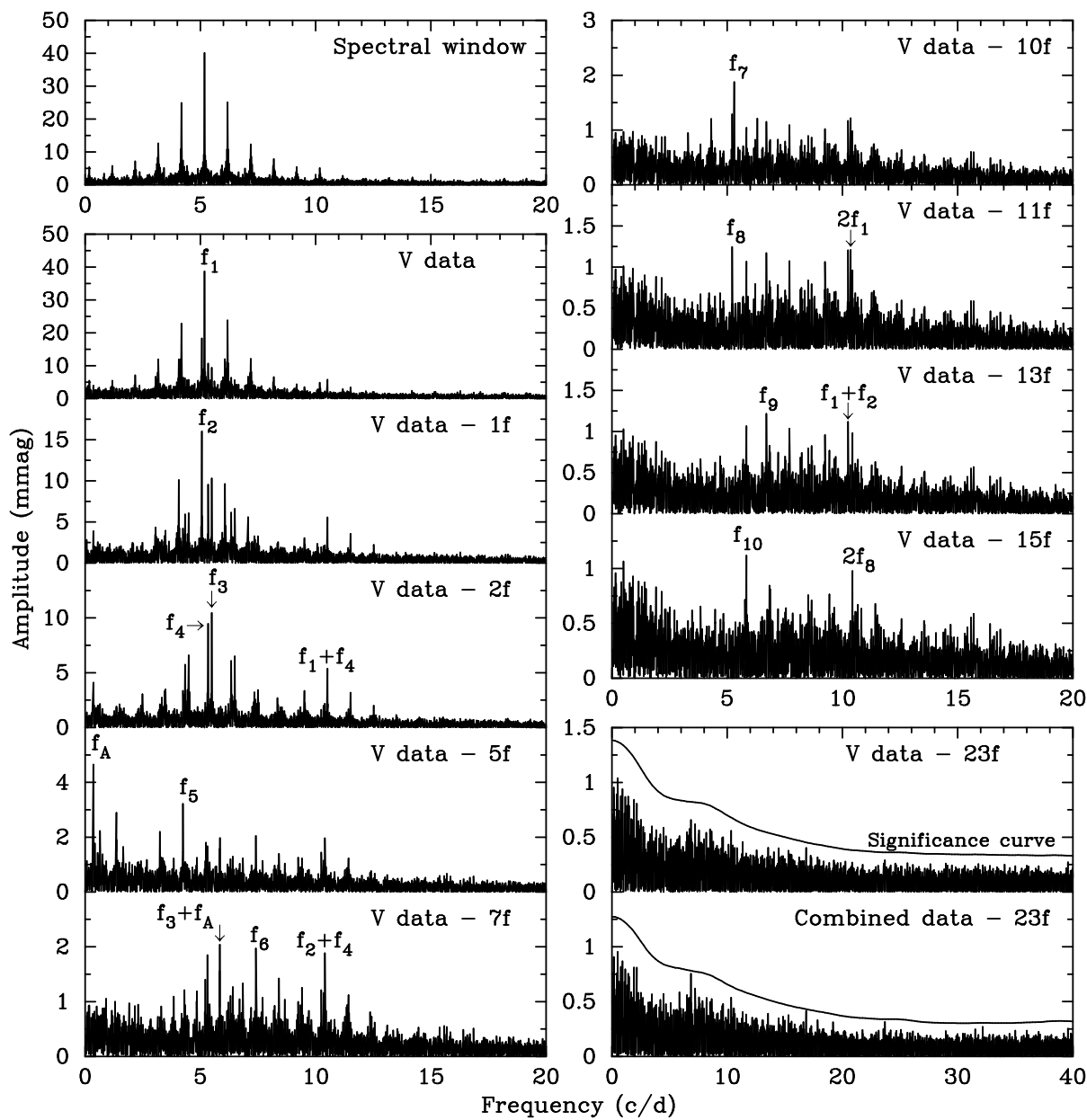}
\caption[]{Amplitude spectra of 12 Lac. The uppermost panel on the
left-hand side shows the spectral window of the data, followed by the
periodogram of the data. Successive prewhitening steps are shown in the
following panels; note their different ordinate scales. After the
detection of 17 periodicities (third panel from bottom, right-hand side),
some low-amplitude combination frequencies were still found in the
residuals. After their prewhitening, the presence of some additional 
periodic signals in the light curves can still be suspected. Note the 
different abscissa scale in the last two panels.}
\end{figure*}

We proceeded by computing the amplitude spectra of the data themselves
(second panel of Fig.\ 2, left-hand side). The signal designated $f_1$
dominates. Prewhitening it by subtracting a synthetic sinusoidal light
curve with a frequency, amplitude and phase that yielded the smallest
possible residual variance, and computing the amplitude spectrum of the
residual light curve, results in the amplitude spectrum in the third panel
of Fig.\ 2, left-hand side.

A second signal ($f_2$) can clearly be seen in this graph; other
variations with similar frequencies also seem present, as proven by
prewhitening $f_1$ and $f_2$ simultaneously (fourth panel of Fig.\ 2,
left-hand side). In this panel we also note the presence of a combination
frequency, the sum of $f_1$ and $f_4$, which was consequently also
prewhitened. We continued this procedure (further panels of Fig.\ 2) until
no significant peaks were left in the residual amplitude spectrum.

We consider an independent peak statistically significant if it exceeds an
amplitude signal-to-noise ratio of 4 in the periodogram; combination
signals must satisfy $S/N>3.5$ to be regarded as significant (see Breger
et al.\ 1993, 1999 for a more in-depth discussion of this criterion). The
noise level was calculated as the average amplitude in a 5 \cd interval
centred on the frequency of interest. Twenty-three statistically
significant sinusoidal variations were found to be necessary to represent
the observed light curves of 12 Lac. We note that the six
lowest-amplitude variations have $S/N \sim 4$, whereas the other signals
all have $S/N > 6$. We also point out that the low-frequency signal $f_A$
is clearly present in the five largest single-site data sets (APT,
Mercator, OSN, SPM, and Lowell) independently and is therefore certainly
real.

Some of the signals we detected are labelled as sums of frequencies found
earlier. All but one of these could be matched with a single pair of
parent modes, and using {\tt Period98} we first verified that the
frequency of the combination would indeed correspond to the sum of the
parent frequencies within the observational errors. Consequently, we fixed
the combination frequencies to the exact predicted value.

We repeated the prewhitening procedure with the $u$ and $v$ data
independently and obtained the same frequencies within the observational 
errors. In particular, we note that we never encountered an alternative 
solution including an alias frequency that fitted the data better. We are 
therefore confident that our frequency determinations are not affected by 
alias ambiguities.

Since the $V$ filter data are most numerous and have the largest time
span, we adopted the frequencies from this data set as our final values
and recomputed the $u$ and $v$ amplitudes with them. Unsurprisingly the
amplitudes did not change significantly from the individually optimised
frequency solution. The final result of our frequency analysis is listed
in Table~2. All the signals were found to be in phase in all filters, 
keeping in mind the observational errors.

\begin{table}
\caption[]{Multifrequency solution for our time-resolved photometry of
12 Lac. Formal error estimates (following Montgomery \& O'Donoghue
1999) for the independent frequencies range from $\pm$ 0.000007 \cd\,for
$f_1$ to $\pm$ 0.00023 \cd\,for $f_{10}$. Formal errors on the amplitudes 
are $\pm$ 0.2 mmag in $u$ and $\pm$ 0.1 mmag in $v$ and $V$. The S/N 
ratio, computed following Breger et al.\ (1993), is for the $V$ filter 
data.} 
\begin{center}
\scriptsize
\begin{tabular}{llcccc}
\hline
ID & Freq. & $u$ Ampl. & $v$ Ampl. & $V$ Ampl. & $S/N$ \\
 & (\cd) & (mmag) & (mmag) & (mmag) & \\
\hline
$f_{1}$ & 5.179034 & 56.4 & 40.7 & 38.1 & 178.6 \\
$f_{2}$ & 5.066346 & 23.3 & 16.7 & 16.0 & 74.6 \\
$f_{3}$ & 5.490167 & 14.2 & 11.7 & 11.1 & 52.4 \\
$f_{4}$ & 5.334357 & 21.9 & 11.6 & 10.0 & 47.3 \\
$f_{5}$ & 4.24062 & 4.4 & 3.7 & 3.6 & 15.8 \\
$f_{A}$ & 0.35529 & 7.2 & 4.8 & 5.0 & 14.4 \\
$f_{6}$ & 7.40705 & 2.8 & 2.1 & 2.0 & 9.7 \\
$f_{7}$ & 5.30912 & 2.7 & 2.3 & 2.0 & 9.5 \\
$f_{8}$ & 5.2162 & 1.3 & 1.3 & 1.3 & 6.2 \\
$f_{9}$ & 6.7023 & 2.2 & 1.6 & 1.3 & 6.3 \\
$f_{10}$ & 5.8341 & 1.8 & 1.2 & 1.3 & 6.1 \\
$f_{1}+f_4$ & 10.513392 & 8.3 & 5.9 & 5.5 & 32.9 \\
$f_{3}+f_A$ & 5.84546 & 2.3 & 1.6 & 1.8 & 8.7 \\
$f_{2}+f_4$ & 10.400704 & 2.3 & 1.7 & 1.7 & 10.3 \\
2$f_{1}$ & 10.358069 & 1.9 & 1.2 & 1.2 & 6.9 \\
$f_{1}+f_2$ & 10.245381 & 1.7 & 1.3 & 1.2 & 6.9 \\
2$f_{8}$ & 10.4324 & 1.5 & 1.4 & 1.0 & 6.1 \\
2$f_{4}$ & 10.668715 & 1.0 & 0.8 & 0.7 & 4.3 \\
$f_{3}+f_4$ & 10.824524 & 0.8 & 0.7 & 0.6 & 3.9 \\
$f_{2}+f_3$ & 10.556514 & 1.1 & 0.8 & 0.7 & 4.0 \\
2$f_1+f_{2}$ & 15.424415 & 0.6 & 0.6 & 0.5 & 4.3 \\
$f_1+f_2+f_4$ & 15.579738 & 1.0 & 0.6 & 0.5 & 4.5 \\
2$f_1+f_4$ & 15.692426 & 1.0 & 0.6 & 0.5 & 4.3 \\
\hline
\end{tabular}
\normalsize
\end{center}
\end{table}

The residuals from this solution were searched for additional candidate
signals that may be intrinsic. We have first investigated the residuals in
the individual filters, then analysed the averaged residuals in the three
main filters (whereby the $u$ data were divided by 1.50 and the $v$ data
were divided by 1.07 to scale possible signal amplitudes to the same level
as in the $V$ data; these scale factors were empirically derived from the
amplitudes of the previously detected signals). The lowest panel on the
right-hand side of Fig.\ 2 contains this final residual amplitude
spectrum. The noise spectrum is not white: a marked increase in amplitude
towards low frequency is clearly visible and additional pulsational
signals may be present.

In particular, two peaks stand out in the combined $uvy$ data prewhitened
by 23 frequencies: a signal at 6.859 \cd and another one at 16.850 \cd,
both with an amplitude signal-to-noise ratio of 3.9. The lower frequency
is very close to the 1 \cd alias of the signal at $f_3+f_A$ and in fact
its amplitude in a multifrequency fit strongly depends on whether
$f_3+f_A$ is assumed to be a combination peak or whether its frequency was
independently optimised. This suggests that $f_3+f_A$ and the signal at
6.859 \cd interact through aliasing. Therefore, the latter cannot be
accepted at this point. As the signal at 16.850 \cd does not correspond
to a combination of the previously identified periodicities (but could be
an alias thereof), we cannot accept it either. These results corroborate
the usefulness of the $S/N>4$ criterion once more.

We note that including residual data in the remaining Geneva filters in
addition did not turn out to be useful because those measurements are not
sufficiently numerous and have a spectral window function too poor for
prewhitening a 23-frequency fit reliably.

\subsection{Some comments on the frequency analysis results}

The residuals between our observed light curves and the 23-frequency fit
are considerably higher than the accuracy of our measurements would imply.
For instance, the residual 12 Lac light curves from Lowell, Fairborn and
San Pedro Martir (our best data) have an rms scatter of 3.7 mmag per
single measurement while the differential comparison-stars magnitudes from
the same sites, after removing the variability of 2 And, show a scatter of
2.1 mmag. The residual rms scatter for the full 12 Lac data set is 
4.0~mmag.

This is not new as far as $\beta$ Cephei stars and even as far as 12  
Lac is concerned. Jerzykiewicz (1978) already noted that the nightly
zeropoints of the data gathered during {\it The International Lacerta
weeks} would vary by up to 0.02 mag. 

In the present $V$ data set, this effect is smaller but clearly present:
the zeropoints vary by up to $\pm$0.008 mag even for the sites with the
most stable equipment. For comparison, the nightly zeropoints of the V
magnitude differences between the comparison stars were found to be
constant within $\pm$ 0.002 mag. Thus, the increase of the noise level
towards low frequencies seen in the last two panels of Fig.\ 2 does not
originate from poor data reductions but is caused by variability of 12
 Lac itself.

We believe that the smaller size of the zeropoint variations in our data
as compared with those in the 1956 data is partly due to the higher
accuracy of our photometry and partly to prewhitening our data with the
low-frequency signal $f_{\rm A}=$ 0.3553 cd${-1}$ before computing the
nightly zeropoints.

It may be that long-term aperiodic variability of $\beta$~Cephei stars is
a common feature, as the same kind of low-frequency noise is present in
the amplitude spectra of other pulsators as well (see Jerzykiewicz et al.\
2005 for $\nu$~Eri or Handler et al.\ 2005 for $\theta$~Oph).

As mentioned in the previous section, we identified a sinusoidal term as a
combination signal if its frequency coincided with the sum or difference
of two or more previously detected oscillations within the observational
errors. This has several implications. Firstly, a combination frequency is
always assumed to have a lower amplitude than the parent modes.
Secondly, a data set with a better frequency resolution than ours may
reveal that some of the combinations we assigned may be chance agreements.
Thirdly, one of our assignments is not unambiguous: the signal identified
with 2$f_4$ can also be matched with $f_1+f_3$ within the accuracy of our
data. This combination may therefore be a mixture of the effect of both
parents.

There are two pairs of close frequencies in Table 2, namely 5.309/5.334
\cd and 5.834/5.845\cd. Their beat periods, 40 and 88 d, respectively, are
well resolved in our data. After prewhitening the two pairs of signals, no
peaks in the vicinity of their frequencies can be found in the residual
amplitude spectrum. Therefore we have no evidence that any of these
variations are due to amplitude or frequency variability of the other
partner in the doublet, and we will assume that all of them are variations
intrinsic to 12 Lac.

\section{Mode identification}

We now attempt to identify the spherical degree $\ell$ of the pulsation
modes by means of the $uvy$ and Geneva passband amplitudes of the
pulsational signals detected in the light curves. These amplitudes are to
be compared with theoretically predicted ones from model computations,
requiring the model parameter space to be constrained first. In other
words, we need to determine the position of 12 Lac in the HR diagram
as a starting point.

The calibrations by Crawford (1978) applied to the mean Str\"omgren colour
indices for 12 Lac listed in the Lausanne Photometric data base ({\tt
http://obswww.unige.ch/gcpd/gcpd.html}) results in $E(b-y)=0.076$. The
Str\"omgren system calibration by Napiwotzki, Sch\"onberner \& Wenske
(1993) gives $T_{\rm eff}=24000\pm1000$~K. With the Geneva colour indices
(again obtained from the Lausanne Photometric data base) of 12 Lac, the
calibrations by K\"unzli et al.\ (1997) provide $T_{\rm
eff}=23500\pm700$~K and log $g=3.4\pm0.4$. The analysis of IUE spectra of
a number of $\beta$ Cephei stars led Niemczura \&
Daszy{\'n}ska-Daszkiewicz (2005) to derive that 12 Lac has $T_{\rm
eff}=23600\pm1100$~K and log $g$=3.65.

To determine the absolute magnitude of 12 Lac, we take advantage of
the fact that it is part of the Lac OB1b association. The distance modulus
of this association was determined with $8.3 \pm 0.3$~mag (Crawford \&
Warren 1976). With $V$=5.25 and the reddening as determined before, we
have $A_V=0.33$ and thus $M_v = -3.4 \pm 0.3$, which we adopt for the
remainder of this work.

According to de Zeeuw et al.\ (1999), the Hipparcos mean distance of Lac
OB1b is $358 \pm 22$ pc, which yields $V-M_v=7.77\pm0.13$ mag. These
authors note that this value is smaller than most previous ones, for
instance by $-0.53\pm0.33$ mag from the one by Crawford \& Warren (1976). 
One possible reason for this result could be a problem with systematic 
errors in Hipparcos parallaxes (Pinsonneault et al.\ 1998).

Summarising the temperature determinations quoted before, we find $T_{\rm
eff}=23700\pm1000$~K for 12 Lac, quite similar to the results by
Dziembowski \& Jerzykiewicz (1999, log $T_{\rm eff} = 4.374 \pm 0.020$).
According to Flower (1996), this effective temperature corresponds to a
bolometric correction of $BC=-2.28\pm0.10$~mag, and therefore $M_{\rm
bol}=-5.7\pm0.4$ mag or $\log L=4.18\pm0.16$. 

The metal abundances derived for the star in the literature show
considerable scatter and we are unable to identify a best value.
Consequently, given the new solar abundances (Asplund et al.\ 2004), we
assume a metallicity of $Z=0.015$ for 12 Lac. We show the position of
the star in a theoretical HR diagram in Fig.\ 3.

\begin{figure}
\includegraphics[width=99mm,viewport=5 00 305 260]{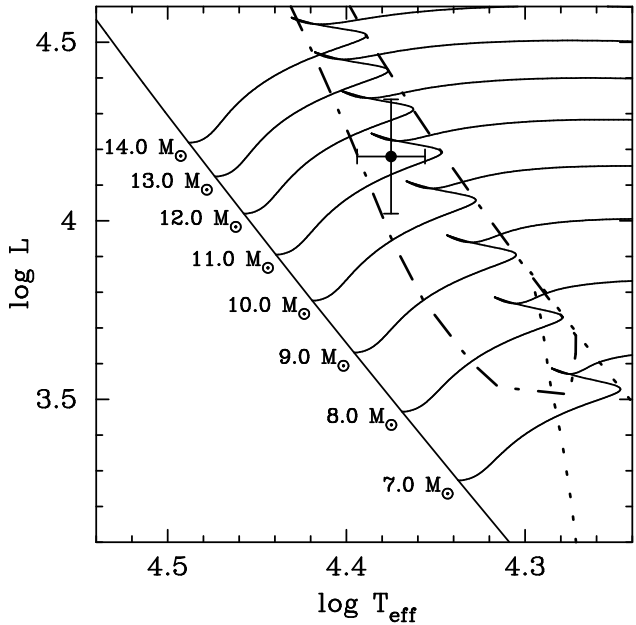}
\caption[]{The position of 12 Lac in the theoretical HR diagram. Some
stellar evolutionary tracks, for a metal abundance of $Z=0.015$, labelled
with their masses (full lines) are included for comparison. We also show
the theoretical borders of the $\beta$~Cephei instability strip
(Pamyatnykh 1999, dashed-dotted line) for $Z=0.015$, and the instability
region for the Slowly Pulsating B (SPB) stars (dotted line).}
\end{figure}

For the purpose of mode identification, we will therefore assume that 12
 Lac is an object of about 11.5 $M_{\odot}$ approaching the end of its
main sequence life. We computed theoretical photometric amplitudes of the
$0 \leq \ell \leq 4$ modes for models with masses between 10.5 and 12.5
$M_{\sun}$ in steps of 0.5 $M_{\sun}$, a temperature range of $4.355 \leq
\log T_{\rm eff} \leq 4.395$ and a metallicity $Z=0.015$. This approach is
similar to that by Balona \& Evers (1999). Theoretical mode frequencies
between 4.0 and 7.6 \cd were considered, except for frequency $f_A$, for
which a theoretical frequency range between 0.3 and 0.4 \cd was examined.
We compare these theoretical photometric amplitude ratios to the observed
ones in Fig.\ 4 for the Str\"omgren $uvy$ filters. As we cannot be certain
whether the signals $f_3+f_A$ and 2$f_8$ are independent modes or not, we
include them in our analysis.

\begin{figure*}
\includegraphics[width=180mm,viewport=00 00 536 521]{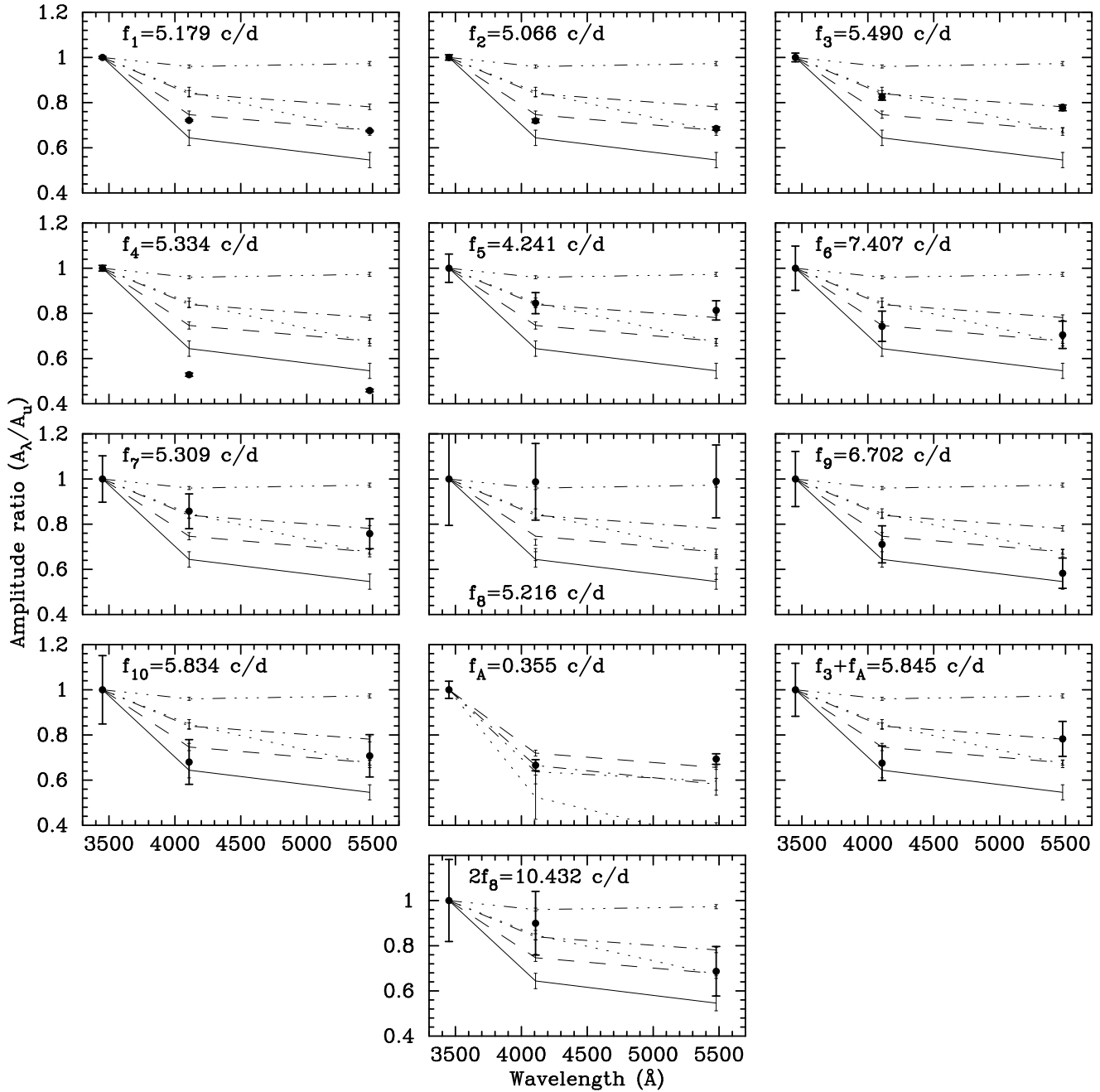}
\caption[]{Mode identifications for 12 Lac from a comparison of
observed and theoretical $uvy$ amplitude ratios, normalised to unity at
$u$. The filled circles with error bars are the observed amplitude ratios.
The full lines are theoretical predictions for radial modes, the dashed
lines for dipole modes, the dashed-dotted lines for quadrupole modes, 
the dotted lines for $\ell=3$ modes, and the dashed-dot-dot-dotted lines 
are for $\ell=4$. The thin error bars denote the uncertainties in the 
theoretical amplitude ratios.}
\end{figure*}

The five pulsations of highest amplitude and the strongest combination
frequency can be easily identified with their spherical degree; we see a
mixture of $\ell=0-2$ modes. For the lower-amplitudes modes, the
identifications become less certain, but at least some possibilities can
be eliminated. For instance, signal $f_9=6.702$~\cd cannot be $\ell=2$ or
4, but it also cannot be $\ell=0$ either because its frequency ratio with
the radial mode $f_4=5.334$~\cd is inconsistent with those for low-order
radial modes. Because of similar arguments with the frequency ratios, we
can also reject $\ell=0$ identifications for $f_{3}+f_A$ and $f_{10}$. The
mode identifications derived this way are listed in Table~3.

\begin{table}
\caption[]{Mode identifications for 12 Lac from our analysis of the 
photometric amplitude ratios.}
\begin{center}
\begin{tabular}{lcc}
\hline
ID & Freq. & $\ell$ \\
 & (\cd) & \\
\hline
$f_{1}$ & 5.179034 & 1\\
$f_{2}$ & 5.066346 & 1\\
$f_{3}$ & 5.490167 & 2\\
$f_{4}$ & 5.334357 & 0\\
$f_{5}$ & 4.24062 & 2\\
$f_{A}$ & 0.35529 & 1, 2 or 4\\
$f_{6}$ & 7.40705 & 1 or 2\\
$f_{7}$ & 5.30912 & 2 or 1 or 3\\
$f_{8}$ & 5.2162 & 4 or 2\\
$f_{9}$ & 6.7023 & 1\\
$f_{10}$ & 5.8341 & 1 or 2\\
$f_{3}+f_{A}$ & 5.84546 & 2 or 1\\
2$f_{8}$ & 10.4324 & 1 or 2 or 3\\
\hline
\end{tabular}
\end{center}
\end{table}

The observed amplitude ratios for the radial mode $f_4=5.334$~\cd are not
well-reproduced by theory. If we attempt to constrain $Z$ from the
observed amplitude ratios, we can only obtain a match if we require $f_4$
to be the fundamental radial mode. In that case, $Z>0.02$. The observed
amplitude ratios cannot be reproduced if $f_4$ was the first or second
overtone (but it has to be a radial mode in any case). We hasten to add
that changing the metal abundance will not significantly affect the
theoretical amplitude ratios for the nonradial modes within the parameter
space under consideration, hence will not affect the $\ell$
identifications for any other mode.

Finally, we also show a comparison of theoretical and observed photometric
amplitude ratios from our Geneva photometry, for the four strongest modes
of 12 Lac (Fig.\ 5). The mode identifications derived from the $uvy$
data are confirmed. For the lower-amplitude modes, such a consistency
check is no longer useful as the observed Geneva amplitude ratios do not
have sufficient accuracy because of the small number of data points
available in these filters.

\begin{figure*}
\includegraphics[width=180mm,viewport=-080 00 456 228]{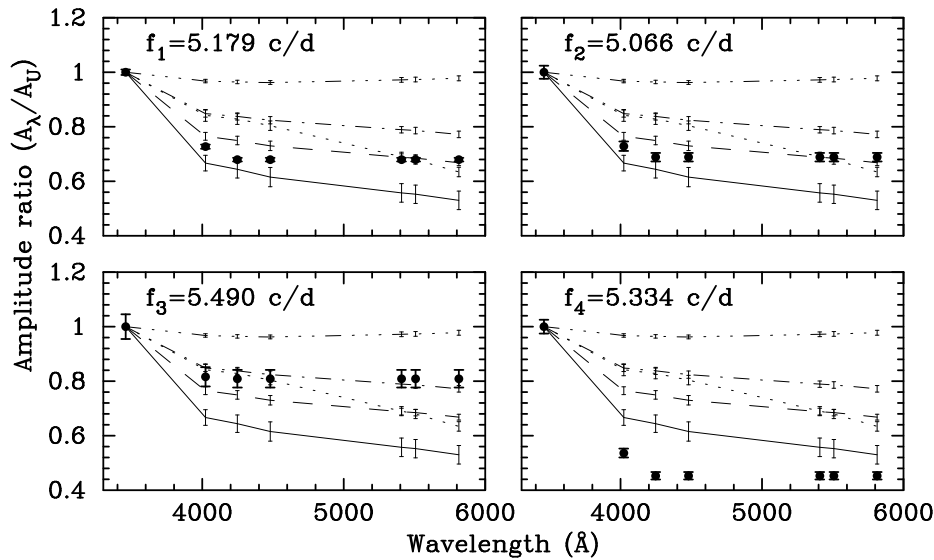}
\caption[]{Identifications for the four strongest modes of 12 Lac
from a comparison of observed and theoretical amplitude ratios in the
Geneva system, normalised to unity at $U$. The filled circles with error
bars are the observed amplitude ratios. The full lines are theoretical
predictions for radial modes, the dashed lines for dipole modes, the
dashed-dotted lines for quadrupole modes, the dotted lines for $\ell=3$ 
modes and the dashed-dot-dot-dotted lines are for $\ell=4$. The thin error 
bars denote the uncertainties in the theoretical amplitude ratios.}
\end{figure*}

\section{Discussion}

\subsection{The independent pulsation modes}

As mentioned in the Introduction, two hypotheses to explain the pulsation
spectrum of 12 Lac seemed promising before our multisite campaign
took place: first, the presence of a rotationally split triplet consisting
of the modes ($f_1, f_3, f_4$) and second, the presence of the fundamental
and first radial overtones (modes $f_5, f_3$). Our mode identification
allows us to judge these hypotheses: neither is correct.

The suspected rotationally split structure consists of modes of $\ell=1,
0, 2$, and the suspected radial modes both turned out to be $\ell=2$. 
Consequently, all previous attempts to understand the pulsation spectrum
of 12 Lac were not correct.

Fortunately, our mode identifications also resulted in the detection of a
radial mode ($f_4$). This will be particularly helpful for the
asteroseismic interpretation of the pulsation spectrum of the star, since
the parameter space in the HR diagram where its seismic model will be
located is greatly reduced. On the other hand, none of the other signals
we detected occurs at a frequency indicative of another radial mode, as
estimated from their frequency ratios to $f_4$.

Two of the strongest modes of 12 Lac are $\ell=1$. It is tempting to
suspect these would be components of a rotationally split multiplet. The
location of the star in the HR diagram as inferred above implies a radius
of $7.0\pm1.8$~R$_{\odot}$. Depending on whether the two modes would be
$|m|=(0,1)$ or $m=(-1,1)$, a rotational velocity of $40\pm10$ or
$20\pm5$~\kms can be inferred, assuming that this splitting reflects that
surface rotation period. The measured $v \sin i$ of 12 Lac is 30~\kms
(Abt, Levato \& Grosso 2002), suggesting that these $\ell=1$ modes are
more likely to be $|m|=(0,1)$. Spectroscopic determinations of the $m$
values of at least the three strongest nonradial modes would be extremely
helpful for the understanding of this star's pulsation spectrum.

The frequency range spanned by the independent modes of 12 Lac
(between 4.241 and 7.407~\cd) is fairly large, and corresponds to three or
four radial overtones. A similarly large range of excited pulsation
frequencies has been found for another $\beta$ Cephei star, $\nu$~Eri
(e.g., see Jerzykiewicz et al.\ 2005), and could not be reproduced by
standard theoretical models (Pamyatnykh, Handler \& Dziembowski 2004;
Ausseloos et al.\ 2004). Consequently, one could suspect that there is a
fundamental problem with our understanding of pulsational driving in the
$\beta$~Cephei stars.

More interesting (and more likely) is the idea that the interior chemical
composition of the star is not homogeneous, a suggestion first brought
forward by Pamyatnykh et al.\ (2004). To drive all the observed modes of
$\nu$~Eri, these authors invoked an ad hoc increase of the abundance of
the iron-group elements by a factor of four in the pulsational driving
region. It may be possible to reconcile pulsational driving of 12 Lac in a
similar way. Theoretical investigations of the question whether or not
diffusion can lead to the required increase of heavy elements in the
driving region while $\beta$ Cephei stars are still on the main sequence
are currently underway (Bourge et al., in preparation).

We also found a low-frequency signal in the light curves of 12 Lac.
Its observed photometric amplitude ratios are consistent with nonradial
pulsation in a $\ell=1, 2$ or 4 mode. On the other hand $f_A$ is the only
signal in this frequency range, and the star's position in the HR diagram
is far away from the SPB star instability strip (Fig.\ 3). Therefore we
cannot be sure about the astrophysical cause of this variation; it could,
for instance, be due to rotational modulation. However, in such a case the
star would rotate rather fast ($v_{\rm rot}\sim120$~\kms, or $v_{\rm
rot}\sim60$~\kms, if the observed frequency was the first harmonic of the
rotation period) which seems unlikely given its measured $v \sin i$ and
the conjectured rotational splitting within the modes of 12 Lac.

\subsection{The combination frequencies - resonant mode coupling?}

We detected several interesting features in the amplitude spectrum of 12
 Lac concerning combination frequencies that prompted us to have a
closer look at these signals. For instance, only combination frequency
sums were found; no frequency differences were detected. If frequency
differences originating from the same parents as the observed frequency
sums had the same amplitudes, three of them should have been detected in
our data. This is similar to what we found for $\nu$ Eri (Handler et al.\
2004).

The single low frequency $f_A$ is also involved in a combination which
resulted in a signal within the range where the intrinsic pulsation mode
frequencies of 12 Lac are located. The amplitude of this combination
signal is unusually high and it is rather unexpected that $f_A$ chose to
combine with $f_3$ only, and not with the highest amplitude mode $f_1$ or
the radial mode $f_4$ which are involved in more combinations than $f_3$.

Quite interestingly as well, the low-amplitude signal $f_8$ was found to
have a harmonic of comparable amplitude, 2$f_8$, that could not be
identified with any other combination of parent frequencies. On the other
hand, 2$f_8$ may be an independent signal that just happens to occur at
the expected frequency of a harmonic. To shed more light on the nature of
$f_8$ and its possible harmonic we have constructed its pulse shape. We
prewhitened the $V$ filter data with all signals but $f_8$ and 2$f_8$, and
then phased the residuals with the parent frequency. The resulting phase
diagram is shown in Fig.\ 6. This graph shows a clear double-wave pulse
shape, which is unlike the ``normal'' phase diagrams of stellar
pulsations, although it should be noted that some stars show similar 
(unphased) light curves (e.g., see Joshi et al.\ 2003).

\begin{figure}
\includegraphics[width=88mm,viewport=-7 7 283 177]{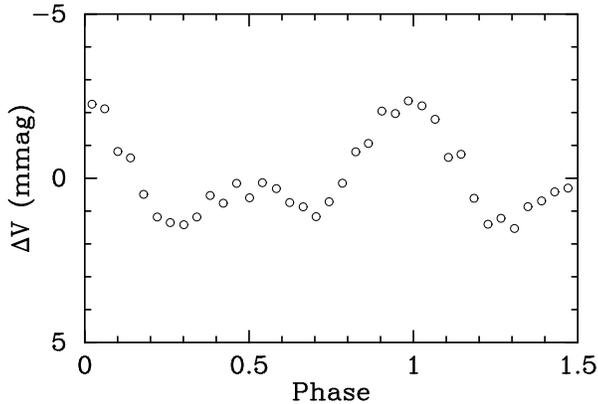}
\caption[]{Phase diagram for the signal at $f_8=5.216$~\cd. The data are 
summed into 25 equally-sized bins for clarity of representation.}
\end{figure}

There are two interpretations for the occurrence of combination
frequencies: light-curve distortions and resonant mode coupling. Under the
first hypothesis, the combinations are caused by the stellar material
being unable to respond totally elastically to the full acceleration due
to pulsation. It should result in combination frequencies whose amplitudes
scale with the product of their parent modes and with their geometrical
cancellation factors. The phases of such combinations relative to those of
their parents should also be similar.

However, such relationships should in general not be followed by modes
excited by resonant mode coupling (Dziembowski 1982). It is the
distinction between the hypotheses of light-curve distortion and resonant
mode coupling as the cause of combination frequencies in the amplitude
spectra of 12 Lac that we are trying to achieve here.

Consequently, we examined the relative amplitudes $A_{ij}/(A_i/A_j)$,
where $A_{ij}$ is the amplitude of the combination frequency and $A_i$ and
$A_j$ are the amplitudes of the parent modes, respectively, and phases
$\phi_{ij}-(\phi_i+\phi_j)$, where $\phi_{ij}$ is the phase of the
combination signal and $\phi_i$ and $\phi_j$ are the phases of the parent
modes of the first-order combination frequency sums with respect to their
parents. Such an analysis has been successfully used by Vuille (2000) and
Vuille \& Brassard (2000) for the pulsating white dwarf star G\,29-38. We
show the relative amplitudes and phases of the combination frequencies in
Fig.\ 7.

\begin{figure}
\includegraphics[width=88mm,viewport=1 7 293 545]{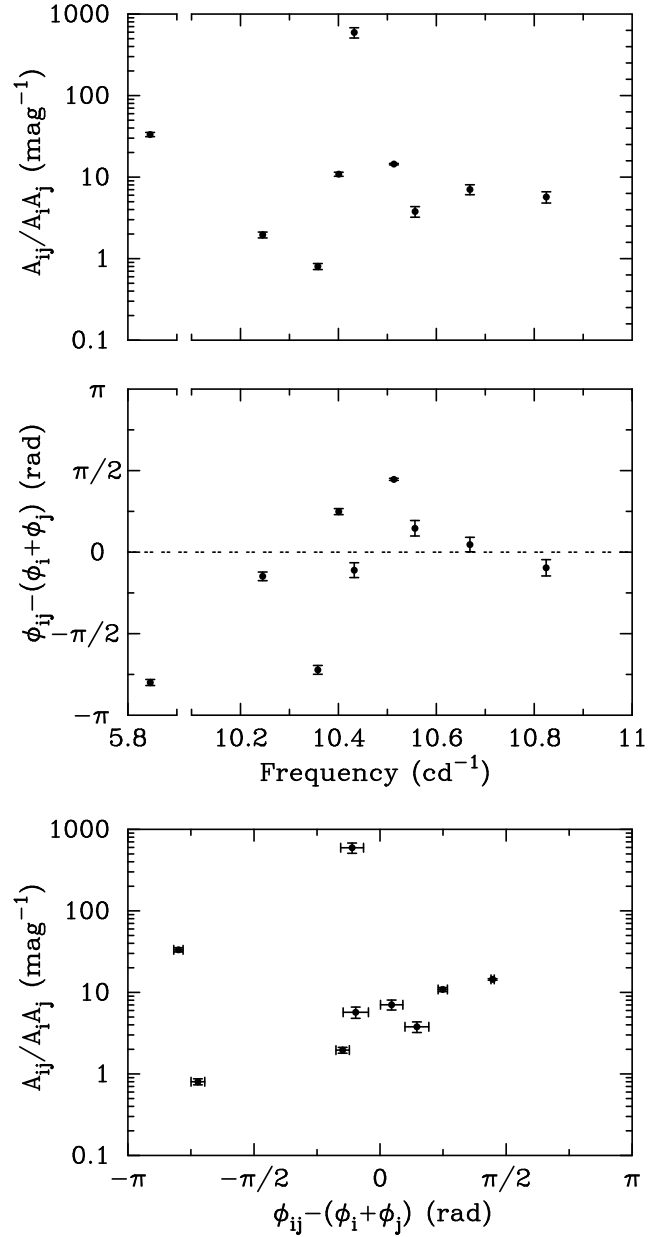}
\caption[]{Relative amplitudes (upper panel) and phases (middle panel) of
the combination frequencies with respect to their parent modes. The 
dotted line in the middle panel connects points of zero phase shift. The 
lower panel shows the relative amplitudes versus relative phase. Note the 
logarithmic scale for the relative amplitudes.}
\end{figure}

Several interesting features can be noted in this graph. First, the
harmonic 2$f_8$ has a relative amplitude almost two orders of magnitude
larger than the other combinations. The second largest relative amplitude
is due to $f_3+f_A$, still a factor of at least three larger than the
others. 

Considering the relative phases, two combinations are again markedly
different from the others: $f_3+f_A$ and 2$f_1$. The relative phase of
2$f_1$ indicates that the pulse shape of this mode has a descending branch
steeper than the rising branch, and that the light maxima are flatter than
the light minima. Given the small amplitude of this harmonic, the effect
is however minimal. In the remaining cases, the relative phases of the
combinations indicate a mixture of different pulse shapes. Some have a
rising branch steeper than the descending branch (positive relative
phase), some behave the opposite way (negative relative phase). Pulse
shapes with descending branches steeper than the rising branches are
unusual for pulsating variables and have to our knowledge only been found
in a few $\delta$ Scuti stars (e.g., see Rodr\I guez et al.\ 1997; Musazzi
et al.\ 1998).

The lowest panel of Fig.\ 7 shows the relative amplitudes with respect to
the relative phases. Again, 2$f_8$ and $f_3+f_A$ stand out, whereas the
other combination frequencies seem to follow a trend. Given that there is
a $2\pi$ ambiguity in the determination of the relative phases, $f_3+f_A$
may actually follow this trend.

In any case, there are two parallel sequences of points along the trend.  
Under the hypothesis of light-curve distortion, this can be understood at
least partly: as the spherical harmonic of these combinations is defined
by the product of the spherical harmonics of its parents, their
photometric amplitudes are subject to geometrical cancellation, including
the effect of inclination. This is consistent with the observations:
besides 2$f_8$ and $f_3+f_A$, the combinations with the highest relative
amplitude involve the radial mode $f_4$, whereas the lower-amplitude
combinations, that should have higher spherical degree and thus suffer
stronger geometrical cancellation, are between nonradial modes. 

Coming back to the nature of the unusual combination frequencies, we have 
one more statement to make: if 2$f_8$ were an independent mode, it would
extend the domain of mode frequencies considerably - and there is already
a problem with driving the frequency range spanned by the other modes!

For all the reasons given above, we can then only explain the occurrence
of 2$f_8$ by nonlinear mode coupling (Dziembowski 1982). In other words,
2$f_8$ is pushed to visible amplitude via a 2:1 resonance with $f_8$. This
would explain the anomalously high amplitude of 2$f_8$, and it would also
give us further clues on pulsational mode identifications. If $f_8$ were
indeed due to an $\ell=4$ mode, it would only be able to interact with
photometrically detectable modes of $\ell=0,2,4$ with the 2:1 resonance
(see Dziembowski 1982). Since we were able to rule out $\ell=0$ or 4 for
2$f_8$ (Fig.\ 4 and Table 3), it can only be a quadrupole mode under this
hypothesis. As the azimuthal order of the resonantly excited oscillation
must be twice the azimuthal order of the exciting mode, we are only left
with the identifications $|m|=0$ or 1 for $f_8$ and $|m|=0$ or 2 for 
2$f_8$. 

A similar situation was found for the $\beta$~Cephei star KK Vel: its
strongest mode has $\ell=4, m=0$. This mode shows a harmonic whose
photometric amplitude ratios are consistent with $\ell=0$ (see the
discussion by Aerts, Waelkens \& de Pauw 1994).

We note in passing that the low photometric amplitude of $f_8$ is not an
argument against it being able to excite another mode via resonant mode
coupling: geometrical cancellation reduces the amplitude of an $\ell=4$
mode by one order of magnitude more than that of an $\ell=2$ mode
(Daszy{\'n}ska-Daszkiewicz et al.\ 2002). Therefore, the intrinsic 
amplitude of $f_8$ would be about a factor of 10 higher than that of 
2$f_8$.

Concerning $f_3+f_A$, we are less certain if it would also be resonantly
excited. Its properties are not as extreme as those of 2$f_8$, although
its amplitude is still unusually high. We would just like to conclude with
another speculation: maybe it is not $f_3+f_A$ that is a resonantly
excited mode. The occurrence of a single self-excited mode at a low
frequency is also not easy to understand. Consequently, it can be
hypothesised that in fact $f_A$ is resonantly excited by $f_3$ and by an
independent mode at $f_3+f_A$.

The behaviour of all the other combination frequencies is consistent with 
the hypothesis of light curve distortion, including $f_1+f_4$ that has 
previously been suggested to be a resonantly excited mode (Aerts 1996).

\section{Conclusions}

The analysis of our extensive photometric observations of the $\beta$
Cephei star 12 Lacertae had a number of interesting surprises to
offer. We added five new independent modes to the five already known,
which may suffice for asteroseismic modelling of the stellar interior. In
particular, one radial mode has been found, which is of great benefit in
restricting the parameter space in which a seismic model is to be located.

Our mode identifications showed that the previously suspected rotationally
split mode triplet was pure coincidence; it actually consists of three
modes of different spherical degree. In this context it is interesting to
note that the $\delta$ Scuti star 1 Mon also exhibits an equally spaced
frequency triplet (Shobbrook \& Stobie 1974), but that the central triplet
component is a radial mode (Balona \& Stobie 1980; Balona et al.\ 2001),
the same situation as found here.

The two pulsation modes of 12 Lac that were suspected to be radial
from their frequency ratios both turned out to be nonradial. These results
are a warning against performing mode identification by just looking for
``suspicious'' structures within the pulsation modes, such as equally
spaced frequencies, and against mode identification by ``magic numbers''
such as the expected frequency ratios of radial modes.

The mode spectrum of 12 Lacertae consists of a mixture of pulsation
modes with spherical degrees between 0 and 4 over a large range of radial
overtones. This is good and bad news for asteroseismology. The good news
is that many modes that sample different regions of the stellar interior
are potentially available. On the other hand, pulsational mode
identification becomes more difficult as the ratio between the number of
observed and theoretically predicted modes is smaller compared to other
$\beta$ Cephei stars. A unique identification of all detected pulsation
modes is therefore not possible from photometry only.

We must therefore put our hopes onto the spectroscopic mode
identifications to follow. Spectroscopic techniques can reveal the
azimuthal order of the modes, making them complementary to the photometric
method. During a spectroscopic analysis it is possible to fix the
spherical degree of the modes to the values following from unique
photometric identifications and to only derive $m$ from the spectroscopy.
Such an approach is probably more robust compared to having to identify
both $\ell$ and $m$ from spectroscopy, and was already successful in the
case of $\theta$~Oph (Briquet et al.\ 2005).

Another interesting result from our study is the detection of a signal 
that could correspond to a linearly stable mode excited by a 2:1 resonance 
via nonlinear mode coupling. This is probably the best case for such a 
mode to be present in a main-sequence pulsator, and it can further be 
tested by deriving more stringent mode identifications from an even larger 
set of photometric measurements. In addition, it may be possible to infer 
the inclination of the star's pulsation axis from the relative amplitudes 
of the ``normal'' combination frequencies.

Finally, we pointed out that the range of excited pulsation frequencies of
12 Lac may be larger than reproducible by standard models. We
therefore suspect that the interior chemical structure of the star is not
homogeneous and that there is probably an increased heavy element
abundance near the pulsational driving zone due to diffusion, as was
already postulated by Pamyatnykh et al.\ (2004) for $\nu$~Eri. Such an
interior compositional stratification could also explain the presence of
$\beta$ Cephei stars in the LMC (Ko{\l}aczkowski et al.\ 2004).

\section*{ACKNOWLEDGEMENTS}

This work has been supported by the Austrian Fonds zur F\"orderung der
wissenschaftlichen Forschung under grant R12-N02. MJ's participation in
the campaign was supported by KBN grant 5P03D01420. ER, PJA and RG are
grateful for the support by the Junta de Andalucía and the Dirección
General de Investigación (DGI) under projects AYA2003-04651 and
ESP2004-03855-C03-01. MJ would also like to acknowledge a generous
allotment of telescope time and the hospitality of Lowell Observatory. JPS
and LP acknowledge the Instituto de Astronomia UNAM and the staff at SPM
Observatory for their continued help. GH wishes to express his thanks to
Lou Boyd and Peter Reegen for their efforts in maintaining and controlling
the Fairborn APTs and to Wojtek Dziembowski for helpful discussions. We
thank the following students of the University of Hawaii for their
assistance during some of the observations: Moses David, Anne Michels, Don
McLemore, Tom Chun, Elizabeth Hart, Erin Brassfield, Robert Knight, David
Plant, Thomas Pegues, Ken Plant, Paul Sherard, Alanna Garay, John Rader
and Jason Barnes.

\bsp


\begin{thebibliography}{99}

\bibitem[]{}Abt H. A., Levato H., Grosso M., 2002, ApJ 573, 359

\bibitem[]{}Adams W. S., 1912, ApJ 35, 163

\bibitem[]{}Aerts C., 1996, A\&A 314, 115

\bibitem[]{}Aerts C., Waelkens C., de Pauw M., 1994, A\&A 286, 136

\bibitem[]{}Aerts C., et al., 2004, A\&A 415, 241

\bibitem[]{}Asplund M., Grevesse N., Sauval A. J., Allende Prieto C., 
Kiselman D., 2004, A\&A 417, 751

\bibitem[]{}Ausseloos M., Scuflaire R., Thoul A., Aerts C., 2004, MNRAS
355, 352

\bibitem[]{}Balona L. A., Stobie R. S., 1980, MNRAS 190, 931

\bibitem[]{}Balona L. A., Evers E. A., 1999, MNRAS 302, 349

\bibitem[]{}Balona L. A., et al., 2001, MNRAS 321, 239

\bibitem[]{}Barning F. J. M., 1963, Bull. Astr. Inst. Netherlands 17, 22

\bibitem[]{}Breger M., et al., 1993, A\&A 271, 482

\bibitem[]{}Breger M., et al., 1999, A\&A 349, 225

\bibitem[]{}Briquet M., Lefever K., Uytterhoeven K., Aerts C., 2005,
MNRAS 362, 619

\bibitem[]{}Crawford D. L., 1978, AJ 83, 48

\bibitem[]{}Crawford D. L., Warren W. H., 1976, PASP 88, 930

\bibitem[]{}Daszy{\'n}ska-Daszkiewicz J., Dziembowski W. A., Pamyatnykh A.
A., Goupil M.-J., 2002, A\&A 392, 151

\bibitem[]{}Dziembowski W. A., 1982, Acta Astr. 32, 147

\bibitem[]{}Dziembowski W. A., Jerzykiewicz M., 1999, A\&A 341, 480

\bibitem[]{}ESA, 1997, The {\it Hipparcos} and Tycho catalogues, ESA
SP-1200

\bibitem[]{}Flower P. J., 1996, ApJ 469, 355

\bibitem[]{}Guthnick P., 1919, AN 208, 219

\bibitem[]{}Handler G. et al., 2004, MNRAS 347, 454

\bibitem[]{}Handler G., Shobbrook R. R., Mokgwetsi T., 2005, MNRAS 362, 
612

\bibitem[]{}de Jager C., 1963, Bull. Astr. Inst. Netherlands 17, 1

\bibitem[]{}Jerzykiewicz M., 1978, Acta Astr. 28, 465

\bibitem[]{}Jerzykiewicz M. et al. 2005, MNRAS 360, 619

\bibitem[]{}Joshi S., et al., 2003, MNRAS 344, 431

\bibitem[]{}Ko{\l}aczkowski Z., et al.\ 2004, in {\it Variable Stars in
the Local Group}, ed. D. W. Kurtz \& K. R. Pollard, ASP Conf.\ Ser.\ 310,
225

\bibitem[]{}K\"unzli M., North P., Kurucz R.\ L., Nicolet B., 1997, A\&AS
122, 51

\bibitem[]{}Ledoux P., 1951, ApJ 114, 373

\bibitem[]{}Mathias P., Aerts C., Gillet D., Waelkens C., 1994, A\&A 289, 
875

\bibitem[]{}Montgomery M. H., O'Donoghue D., 1999, Delta Scuti Star
Newsletter 13, 28 (University of Vienna)

\bibitem[]{}Musazzi F., Poretti E., Covino S., Arellano Ferro A., 1998, 
PASP 110, 1156

\bibitem[]{}Napiwotzki R., Sch\"onberner D., Wenske V., 1993, A\&A 268,
653

\bibitem[]{}Niemczura E., Daszy{\'n}ska-Daszkiewicz J., 2005, A\&A 433,
659

\bibitem[]{}Pamyatnykh A. A., 1999, Acta Astr. 49, 119

\bibitem[]{}Pamyatnykh A. A., Handler G., Dziembowski W. A., 2004, MNRAS
350, 1022

\bibitem[]{}Pinsonneault M.\ H., Stauffer J., Soderblom D.\ R., King J.\
R., Hanson R.\ B., 1998, ApJ 504, 170

\bibitem[]{}Rodr\I guez E., Gonz\'alez-Bedolla S. F., Rolland A., Costa
V., L\'opez-Gonz\'alez M.\ J., L\'opez de Coca P., 1997, A\&A 328,~235

\bibitem[]{}Sareyan J. P., Chauville J., Chapellier E., Alvarez M., 1997,
A\&A 321, 145

\bibitem[]{}Shobbrook R. R., Stobie R. S., 1974, MNRAS 169, 643

\bibitem[]{}Smith M. A., 1980, ApJ 240, 149

\bibitem[]{}Sperl M., 1998, Master's Thesis, University of Vienna

\bibitem[]{}Stankov A., Handler G., 2005, ApJS 158, 193

\bibitem[]{}Stebbins J., 1917, Pop. Astr. 25, 657

\bibitem[]{}Vuille F., 2000, MNRAS 313, 179

\bibitem[]{}Vuille F., Brassard P., 2000, MNRAS 313, 185

\bibitem[]{}de Zeeuw P.\ T., Hoogerwerf R., de Bruijne J.\ H.\ J., Brown
A.\ G.\ A., Blaauw A., 1999, AJ 117, 354

\end{thebibliography}
\end{document}